\documentclass[letter]{aa} 

\usepackage{txfonts}
\usepackage{graphicx}
\usepackage{natbib}
\usepackage{mathrsfs}
\usepackage{url}

\begin{document}

\title{HH~114~MMS: a new chemically active outflow
\thanks{Based on observations carried out with the IRAM 30m Telescope. IRAM is supported by INSU/CNRS (France), MPG (Germany) and IGN (Spain)}}

\author{M. Tafalla \inst{1}
\and
A. Hacar \inst{1,2}
}

\institute{Observatorio Astron\'omico Nacional (IGN), Alfonso XII 3, 
E-28014 Madrid, Spain
\email{m.tafalla@oan.es}
\and
Institute for Astrophysics, University of Vienna,
T\"urkenschanzstrasse 17, A-1180 Vienna, Austria\\
\email{alvaro.hacar@univie.ac.at}
}

\date{}

\abstract
{A small group of bipolar protostellar outflows display strong emission from shock-tracer
molecules such as SiO and CH$_3$OH, and are generally referred to as 
``chemically active.'' The best-studied outflow from this group
is the one in L~1157.}
{We study the molecular emission from the bipolar outflow powered by the
very young stellar object HH~114~MMS and compare its chemical composition
with that of the L~1157 outflow.}
{We have used the IRAM 30m radio telescope to observe a number of 
transitions from CO, SiO, CH$_3$OH, SO, CS, HCN, and HCO$^+$ toward
the HH~114~MMS outflow. The observations consist of maps and 
a two-position molecular survey. }
{The HH~114~MMS outflow presents strong emission from a number of 
shock-tracer molecules that dominate the appearance of the
maps around the central source. The
abundance of these molecules  is comparable to the abundance  in L~1157.}
{The outflow from HH~114~MMS is a spectacular new case of 
a chemically active outflow.}


\maketitle
%

\section{Introduction}
\label{intro}

Bipolar outflows from young stellar objects (YSOs)
are routinely identified by their mechanical impact
on the surrounding cloud,
which produces
characteristic lobes of blue- and red-shifted material.
A small group of outflows also show evidence 
of chemical impact in the form of enhanced 
abundance of shock-sensitive species such as SiO and CH$_3$OH.
These outflows are often referred to as ``chemically
active,'' and their study offers a unique opportunity 
to investigate both ambient gas acceleration and
the complex reactions of shock chemistry \citep{taf11}.

Here we report the serendipitous identification
of  a chemically active outflow driven by HH~114~MMS.
HH~114~MMS is located in the L~1589 cloud, at an 
estimated distance of 450~pc \citep{mat08}.
It was discovered at mm-wavelengths by \citet{chi97},
and was later mapped at cm wavelengths by \citet{rod96}.
Both HH~114~MMS and the neighboring source IRAS~05155+0707
lie near the center 
of the 2.6-pc long 
chain of Herbig-Haro (HH) objects HH~114/HH~115 \citep{rei97},
although the relation between the two YSOs and the HH chain remains unclear.
Each YSO drives a separate CO outflow, and there is
no evidence for any interaction between the two. 
As shown by \citet{lee02}, the blue lobe of 
the HH~114~MMS outflow approximately coincides with HH~114D, but the 
large-scale orientation of the HH~114/HH~115 system differs 
significantly from the direction of the two CO outflows. While
it is possible that IRAS~05155+0707 drives the large-scale HH
chain (if its initially E-W direction changes to SE-NW), it is
unlikely that the HH~114~MMS outflow plays any role in the
large-scale HH~114/HH~115 system, since that would require
a change of outflow direction close to 90 degrees.

HH~114~MMS is a highly embedded object. 
\citet{chi97} suggested it is a class~0
object \citep{and93}, and determined an upper limit
to its luminosity of about 25~L$_\odot$. Further evidence for
its extreme youth 
comes from the high collimation of its molecular outflow
shown by the interferometric maps of \citet{arc06}.

\section{Observations}
\label{obs}

The HH~114~MMS bipolar outflow was observed in the CO(2--1) line with
the IRAM 30m telescope during December 2008
using the dual-polarization $3\times 3$ HERA array operating in 
wobbler-switching mode with a throw of $240''$ and a frequency of
0.5~Hz. The signal from HERA was sent to the
VESPA autocorrelator, which was configured 
to produce spectra with a resolution of 1.25~MHz (1.6~km~s$^{-1}$) 
over a 280~MHz (365~km~s$^{-1}$) passband. These spectra 
were later processed with the {\tt CLASS} 
software\footnote{\url{http://www.iram.fr/IRAMFR/GILDAS}}
and converted to the main beam brightness
temperature scale.

\begin{figure*}
\centering
\resizebox{\hsize}{!}{\includegraphics{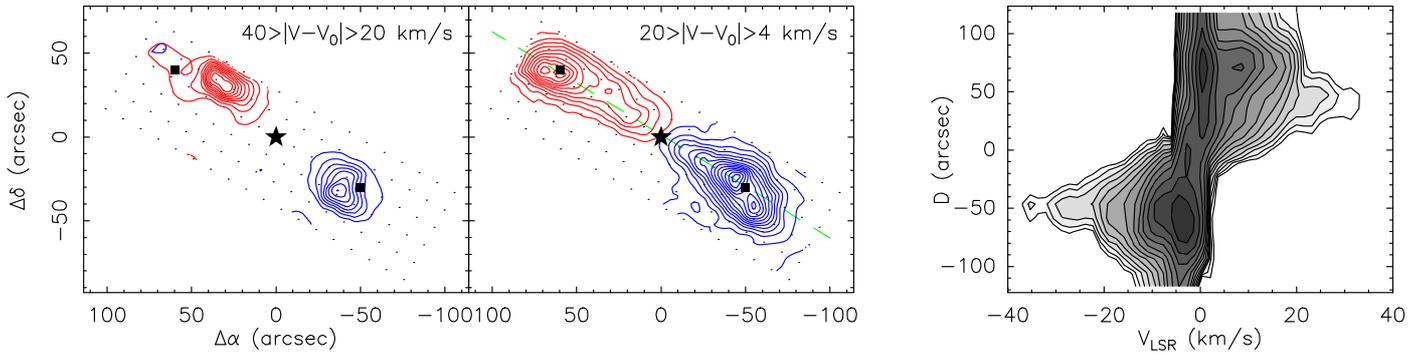}}
\caption{Structure of the HH~114~MMS outflow from CO(2--1) data.
{\em Left: } maps of intensity integrated in 
velocity ranges symmetrically separated from the ambient cloud
velocity ($V_0 = -0.5$~km~s$^{-1}$ from N$_2$H$^+$ data). 
First contour and contour
interval are 1~K~km~s$^{-1}$ in the high-velocity map
and 5~K~km~s$^{-1}$ in the low-velocity map.
The star symbol at the offset origin 
represents the position of HH~114~MMS
($\alpha(\mathrm{J}2000)=05^h18^m15\fs2,$
$\delta(\mathrm{J}2000)=+7^\circ12'2''$), the
two squares are the positions where
the line survey was carried out, and the dots are observed positions.
The dashed green line is the direction of the position-velocity 
diagram shown in the next panel.
{\em Right: } position-velocity diagram 
along the outflow axis. To emphasize the low-level
outflow emission, the contours are spaced by factors of 
square root of 2
starting at 0.1~K.
\label{co-maps}}
\end{figure*}

Additional IRAM 30m observations of the HH~114~MMS outflow were made in 
February 2011. 
An initial survey for selected molecular lines  in 
the 1, 2, and 3~mm wavelength bands was carried out toward
several positions.
In view of the bright lines of shock-tracer species,
additional positions were observed 
to map the full extent of the emission. 
These shock-tracer observations were complemented 
with limited maps of the quiescent-gas tracers 
N$_2$H$^+$(1--0) and C$^{18}$O(2--1).
All observations were made with the dual-band, dual-polarization, 
single-pixel
EMIR receiver. For the simultaneous 
N$_2$H$^+$ and C$^{18}$O observations, 
we used frequency-switching mode to ensure that the spectra 
were free from ambient-cloud contamination. For the
rest of the observations, we used the same wobbler-switching mode used
in the CO(2--1) observations.
As spectrometers, we used the FTS and VESPA instruments, which,
depending
on the line, provided velocity resolutions between 0.06 and 
0.7~km~s$^{-1}$. The resulting spectra were processed with CLASS
as done with the CO(2--1) spectra. 

\section{Physical properties of the HH~114~MMS outflow}
\label{phys}

Fig.~\ref{co-maps} summarizes the main features of the
CO(2--1) data.
The two  maps in the left and middle panels show that
the CO emission has a markedly bipolar
distribution with respect to HH~114~MMS,
with the redshifted emission  toward the northeast
and the blueshifted emission  toward the southwest.
This distribution agrees with that found
by \citet{arc06} from interferometric observations
of the central arcminute around the YSO,
and with the distribution
of blue CO emission mapped by \citet{lee02},
also with an interferometer (and restricted
to the southern blue lobe).
Our more extended CO(2--1) maps show that the HH~114~MMS 
outflow has a high degree of collimation at all
velocities, and that its emission extends for at least $200''$ (0.4~pc
for the assumed distance of 450~pc).

To illustrate the velocity field of the gas, we present 
a position-velocity diagram along the outflow
axis in the rightmost panel of Fig.~\ref{co-maps}.
This diagram shows that both outflow lobes 
follow an approximate 
linear increase of the terminal velocity with distance from
the central source. 
The ``Hubble-law'' velocity pattern, however, is only
approximate, since the
fastest velocities are not reached at 
the end of the lobes but at an intermediate distance. 
This can also be seen in the velocity maps of Fig.~\ref{co-maps}, 
which show that the gas faster than 20~km~s$^{-1}$ forms two 
bright spots about $50''$ from the YSO.
The origin of this velocity pattern is unclear, and 
the 1-beam sampling of our data is too coarse
to allow further investigation. 
The blue-red
symmetry of the pattern, however, suggests that it is intrinsic to the outflow,
and not the result of an interaction with the ambient cloud.
While unusual, the pattern seems not unique to HH~114~MMS. Similar
cases, together with their difficult interpretation in terms of 
existing outflow models, have been discussed by \citet{che95}.

We complete our analysis 
estimating the energetics of the moving gas traced 
by CO(2--1). 
We follow standard practice
and do not  correct for the uncertain inclination
angle of the outflow. The lack of overlap between the
blue and red lobes, together with the lack of
color mixing inside each lobe, make HH~114~MMS 
a Case 2 outflow in the classification of \citet{cab86}, 
suggesting an inclination angle between
$15^\circ$ and $75^\circ$ (we estimate an outflow
opening angle $\theta_{\mathrm{max}} = 15^\circ$ 
from the CO maps).
As a result, the inclination correction
to the velocity
lies between 4\% and a factor of 4,
and this makes our energetics estimate a strict lower
limit.
To avoid contamination from ambient-cloud material,
we also  limit our analysis to
emission faster than 3~km~s$^{-1}$.
We also assume that
 the CO(2--1) emission is optically thin and in 
local thermodynamic equilibrium (LTE), with an
excitation temperature of 25~K 
(from the CO(3--2)/CO(2--1) ratio,
Sect.~\ref{abu-sec}), and use a 
CO abundance of $8.5 \times 10^{-5}$
\citep{fre82}. 

With these  assumptions, 
we estimate a total outflow mass of 0.15~M$_\odot$,
a momentum of 1.3~M$_\odot$~km~s$^{-1}$, and
a kinetic energy of $3.5 \times 10^{44}$~erg.
These values are between four times less than
(mass and momentum) and equal to (energy)
those estimated for L~1157 by \citet{bac01},
who do not correct for inclination in
this also Case 2 outflow.

\section{Maps of dense-gas tracers}
\label{molec}

\begin{figure*}
\centering
\resizebox{\hsize}{!}{\includegraphics{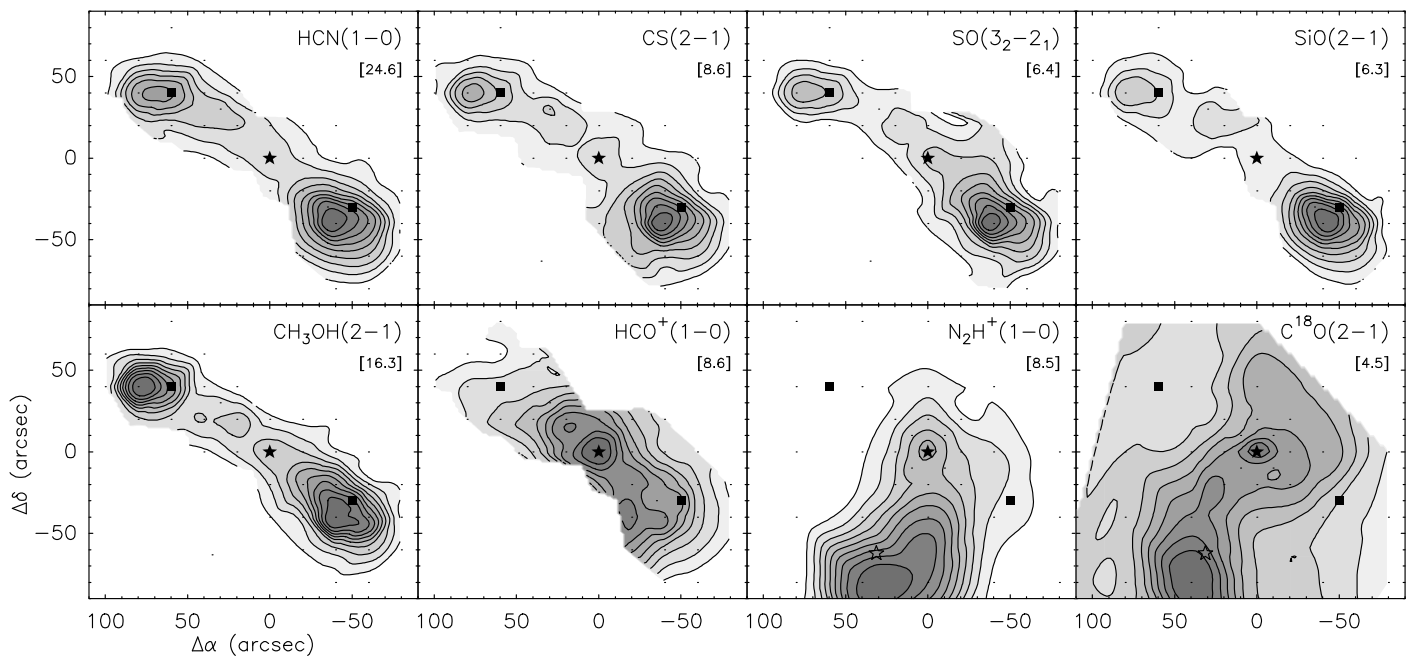}}
\caption{Emission from dense-gas tracers in the 
vicinity of HH~114~MMS.
The maps represent emission integrated over the complete velocity range
of detection. Contours are at 10, 20, ... 90\% of the map maximum,
whose value in K~km~s$^{-1}$ is given within brackets under the name
of the transition.
Central positions, dots, and solid star as in Fig.~\ref{co-maps}.
The open star in the last two panels represents IRAS 05155+0707.
\label{shock-maps}}
\end{figure*}

While the CO maps of the HH~114~MMS outflow
are not qualitatively different from 
those of other young outflows (e.g., 
\citealt{lee02,arc06}), the maps of
the high-density tracers are truly remarkable.
Fig.~\ref{shock-maps} shows that 
the emission of 
dense gas tracers such as 
HCN, CS, SO, CH$_3$OH, or SiO
is not concentrated toward the
central source, 
but presents two bright peaks toward the outflow lobes.
HCO$^+$, while centrally concentrated, 
also shows an elongation in the direction
of the outflow, and 
only N$_2$H$^+$, and C$^{18}$O (the last one
not strictly a high-density tracer) present 
compact emission peaks towards HH~114~MMS.
(The southern  
N$_2$H$^+$ and C$^{18}$O peak coincides
with the vicinity of 
IRAS~05155+0707.)

The similarity between the emission of the dense-gas 
tracers and the CO outflow is not limited to 
the integrated maps of Fig.~\ref{shock-maps}. 
Velocity maps using the intervals 
of Fig.~\ref{co-maps}
show the same blue-red bipolar distribution and 
the same offset between high- and 
low-velocity peaks.
Individual spectra, like those in  Fig.~\ref{spec},
show prominent wings of clear outflow origin.

Bipolar outflows are known to sometimes enhance the 
emission of certain molecular species, especially
SiO and CH$_3$OH, due to the disruption of dust
grains and the ices that coat them \citep{van98}.
Few outflows, however, display the 
dramatic 
impact seen in the emission of other tracers
associated with dense core gas, such as
CS and HCN. These species are routinely used
to trace the innermost environment around YSOs
(e.g., \citealt{afo98,lau98}), and are not
expected to represent outflow gas.
As Fig.~\ref{shock-maps} shows, this is
not the case in HH~114~MMS, where the
core emission is dwarfed by the outflow
contribution.

As mentioned before, only a small group of 
outflows present the
prominent emission from dense-gas tracers found 
towards HH~114~MMS. 
The best-known member of this group is
the L~1157 outflow, which has long been
recognized for presenting unusually strong 
emission from shock-tracer molecules,
such as SiO \citep{mik92}.
The molecular maps 
of \citet{bac01} (their Fig.~6)
show that most dense-gas tracers in L~1157
(including HCN, CS, 
SO, CH$_3$OH, and SiO) present a remarkable 
pattern of outflow-dominated emission, similar
to that shown in Fig.~\ref{shock-maps}
for HH~114~MMS. 

The dramatic effect of the L~1157 outflow in
the chemistry of its gas 
led \citet{bac01} to coin the term ``chemically active''
to refer to this outflow and others 
with similar properties,
such as BHR~71 \citep{bou97,gar98}.
The maps of Fig.~\ref{shock-maps} show now
that the 
term can also be applied to the HH~114~MMS
outflow, which appears as
one of the finest examples of this class,
thanks to its clean pattern of a single
bright emission peak near the end of two
lobes of similar size.

\section{Column density and abundance estimates}
\label{abu-sec}

\begin{figure}
\centering
\resizebox{\hsize}{!}{\includegraphics{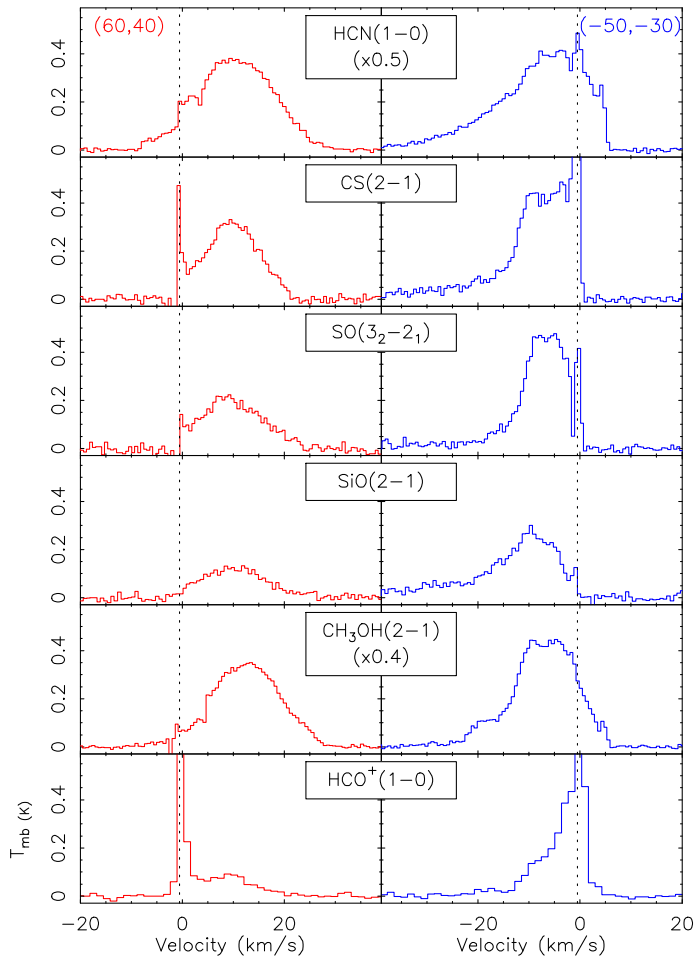}}
\caption{Sample spectra from the line survey carried out
toward two outflow positions. The offset of each
position is indicated in the top panel (in arcseconds
and referred to the map center of Fig.~\ref{co-maps}).
The vertical dashed line indicates the velocity of the
ambient cloud.
\label{spec}}
\end{figure}

To further characterize the chemistry of the 
HH~114~MMS outflow,
we convert the observed intensities
into molecular column densities. 
For this, we use the results of a small survey of 
transitions at 1, 2, and 3~mm wavelength toward 
($60''$, $40''$) and ($-50''$, $-30''$).
Representative spectra from this survey are 
shown in Fig.~\ref{spec}. The full list of 
detections include
CO(2--1), CO(3--2), CS(2--1), CS(3--2), 
SiO(2--1), SiO(3--2), SO(3$_2$--2$_1$), SO(4$_3$--3$_2$),
CH$_3$OH(2$_k$--1$_k$), CH$_3$OH(3$_k$--2$_k$), 
HCN(1--0), HCO$^+$(1--0), and H$_2$CO(2$_{02}$--1$_{01}$).

At the time of the observations, and based on our
CO maps, the two survey positions were thought
to represent the brightest peaks
of each outflow lobe.
Later mapping in multiple molecules, however,
showed that these positions were 
offset from the true emission peaks by
$\sim 15''$.
While these small offsets are not critical for the abundance results,
which seem almost position-independent, they may affect 
the comparison of the HH~114~MMS outflow
intensities with 
those of other objects.

As a first step in our column density estimate
we chose for each survey
position a 20~km~s$^{-1}$-wide interval of outflow velocities
(see values in Table~\ref{abu}). 
The molecular emission in this interval  was 
analyzed with an LTE method similar to that 
presented in \citet{taf10}, although no dilution factors
were applied to the data because the observations were not
centered on the emission peak. 
For the CO emission, we used data from
5-point crosses with a $5''$ sampling
to convolve the CO(3--2)
emission to the $12''$ resolution of the CO(2--1) data.
These two intensities were used to derive CO
excitation temperatures between 25 and 30~K
and the column density values reported in Table~\ref{abu}.
For the dense-gas tracers with two observed
transitions (CS, SiO, SO, and CH$_3$OH), we used
population diagrams to determine simultaneously
column densities and excitation temperatures, which
range from about 4 to 8~K and indicate that
the molecules are subthermally excited.
Finally, for
the three species with only one observed transition
(HCN, HCO$^+$, and para-H$_2$CO), we 
assumed an excitation temperature of 6~K
(mean value of CS, SiO, SO, and CH$_3$OH),
and used the integrated 
intensities to estimate  molecular column densities.

Using the previous column densities, we determined
for each species an
abundance with respect to CO, and
compared this value with the typical undepleted value
in a dense core as discussed in \citet{taf10}.
The results of this estimate 
are presented in Table~\ref{abu} in terms of
the enhancement factor $f_{\mathrm{enh}}$, which is the 
ratio between the outflow and the core CO-normalized abundances.
As can be seen, the $f_{\mathrm{enh}}$ factors
in the two outflow positions 
agree with each other within
a factor of 2 for all observed species.
This agreement, which is probably at the level of our 
measurement uncertainty, suggests that the 
two outflow lobes are chemically very similar. 

The results in
Table~\ref{abu} also show that the enhancement factors 
of the different species span a wide range of values.
The smallest factor is that of HCO$^+$, and is
consistent with this molecule having 
little or no enhancement in the outflow gas.
This result is consistent with the map in
Fig.~\ref{shock-maps}, which shows no emission peak
for this molecule toward neither of the outflow lobes
(although the emission is elongated in the outflow direction).
All other species in Table~\ref{abu}
present significant abundance enhancements, and
range from about one order of magnitude in the case 
of HCN and CS to more than three orders of magnitude
in the case of CH$_3$OH and SiO.

While large, the enhancement factors in Table~\ref{abu}
are comparable to those in other young outflows.
This can be seen by comparing the values in the table
with those presented by \citet{taf10}, who
measured enhancement factors in the
L~1448 and IRAS~04166 outflows and re-evaluated the L~1157-B1
estimates previously made by \citet{bac97}.
The values for these three outflows are in good agreement with our 
HH~114~MMS estimates, and for
example, the geometrical mean of the $f_{\mathrm{enh}}$
estimates in the two HH~114~MMS survey positions 
agrees with those in  L~1157-B1
better than a factor of 3
for all molecular species. 
Such an agreement in enhancement factors
confirms the idea that HH~114~MMS must be
a {\em bona fide} chemically active outflow.

Although the abundances with respect to CO in the 
HH~114~MMS outflow are comparable to those in L~1157, 
the integrated intensities 
towards its brightest molecular peak 
are lower than those of L~1157-B1 by an average factor
of 2.5. 
This suggests that the HH~114~MMS outflow may not have
accumulated as much enriched material as
L~1157, although beam dilution may still affect the 
HH~114~MMS intensities. 
The more southern location of HH~114~MMS,
on the other hand, makes this outflow 
a good candidate for ALMA observations, something
impossible with  L~1157 due to its high declination.
The newly-identified chemically active HH~114~MMS 
outflow therefore has significant
potential to shed new light
on the still-mysterious nature of this outflow class.

\begin{table}
\caption[]{Survey column densities and abundance enhancements.}
\label{abu}
\centering
\begin{tabular}{l c c c c}
\hline
\noalign{\smallskip}
& \multicolumn{2}{c}{($60''$, $40''$)\tablefootmark{a}} &
\multicolumn{2}{c}{($-50''$, $-30''$)\tablefootmark{b}} \\
\noalign{\smallskip}
\mbox{MOL}  & $N$(cm$^{-2})$  & 
$f_\mathrm{enh}\tablefootmark{c}$
& $N$(cm$^{-2})$  & $f_\mathrm{enh}\tablefootmark{c}$ \\
\noalign{\smallskip}
\hline
\noalign{\smallskip}
CO & $1.8\; 10^{16}$ & - & $2.1\; 10^{16}$ & - \\
HCO$^+$ & $1.0\; 10^{12}$  & 1.1 & $1.2\; 10^{12}$ & 1.2 \\
HCN &  $2.1\; 10^{13}$ & 11  & $2.1\; 10^{13}$ & 7.2\\
CS & $1.8\; 10^{13}$  & 14  & $2.8\; 10^{13}$ & 19 \\
p-H$_2$CO & $8.2\; 10^{12}$ & 68  & $9.5\; 10^{12}$  & 65 \\
SO & $3.4\; 10^{13}$  & 290 & $4.4\; 10^{13}$ & 320 \\
CH$_3$OH & $4.6\; 10^{14}$  & $1.8\; 10^3$  & $3.2\; 10^{14}$ & $1.1\; 10^3$ \\
SiO & $5.5\; 10^{12}$ & $5.3\; 10^3$  & $1.4\; 10^{13}$  & $1.1\; 10^4$ \\
\hline
\end{tabular}
\tablefoot{
\tablefoottext{a}{Quantities measured in the $V_{\mathrm{LSR}}$
range 4 to 24~km~s$^{-1}$.}
\tablefoottext{b}{Quantities measured in the $V_{\mathrm{LSR}}$
range -26 to -6~km~s$^{-1}$.}
\tablefoottext{c}{Abundance enhancement factor over dense core value as in
\citet{taf10}.  }}
\end{table}

\begin{acknowledgements}
We thank the observers of the HERA pool for obtaining the
CO(2--1) data, and the IRAM staff for help during the
EMIR observations. The anonymous referee and Malcolm Walmsley 
provided a number of useful comments.
MT acknowledges support from the MICINN program
CONSOLIDER INGENIO 2010, grant ``Molecular Astrophysics:
The Herschel and ALMA era - ASTROMOL'' (ref.: CSD2009-00038).
This publication is also supported by the
 Austrian Science Fund (FWF).
This research has made use of NASA's Astrophysics Data System
Bibliographic Services together with the SIMBAD database and
the VizieR catalogue access tool
operated at CDS, Strasbourg, France.
\end{acknowledgements}

\end{document}